# ENHANCING INTEROPERABILITY AMONG HEALTH INFORMATION SYSTEMS IN LOW AND MIDDLE-INCOME COUNTRIES: A REVIEW OF CHALLENGES AND STRATEGIES

Prabath Jayatissa and Roshan Hewapathirane

University of Colombo

## ABSTRACT

*The review article aims to provide an overview of the challenges and strategies for enhancing interoperability among health information systems in low and middle-income countries (LMICs). Achieving interoperability in LMICs presents unique challenges due to various factors, such as limited resources, fragmented health information systems, and diverse health IT infrastructure. The methodology involves conducting a comprehensive literature review, synthesising findings, identifying challenges and strategies, analysing and interpreting results, and writing and finalising the article. The article highlights that the interoperability challenges include a lack of standardisation, fragmented systems, limited resources, and data privacy concerns. The article proposes strategies to enhance interoperability in LMICs, such as standardisation of data formats and protocols, consolidation of health information systems, investment in health IT infrastructure, and capacity building of health IT professionals in LMICs. The article aims to provide insights into the current state and potential strategies for enhancing interoperability among health information systems in LMICs, intending to improve healthcare delivery and outcomes in these*

## 1. INTRODUCTION

Health information systems (HISs) are critical in supporting the delivery of quality healthcare services, improving patient outcomes, and informing evidence-based decision-making. HIS faces unique challenges in low- and middle-income countries (LMICs) due to limited resources, infrastructure, and capacity. One key challenge is the lack of interoperability, which refers to the ability of different HISs to exchange and use health information effectively.

Interoperability is crucial for achieving seamless data exchange, coordination of care, and integration of health services. It enables healthcare providers to access and share patient information across different systems, facilitating timely and informed decision-making. However, achieving interoperability in LMICs is complex and poses several barriers, including technical, organisational, and financial challenges.

This review article aims to highlight the challenges and strategies for enhancing interoperability among HIS in LMICs. We will explore the current state of interoperability in LMICs, identify the challenges faced by these countries, and discuss the strategies that can be employed to overcome these challenges. We will also highlight the importance of standardization, consolidation, investment, and capacity building as key strategies for improving interoperability in LMICs. By addressing these challenges and implementing appropriate strategies, LMICs can unlock the full potential of HIS to improve healthcare outcomes and achieve better health for their populations.

## 2. METHODOLOGY

The review article focuses on the issue of interoperability among health information systems in low and middle-income countries (LMICs). The methodology employed involves a rigorous process that includes several key steps. First, a comprehensive literature review encompasses relevant research articles, reports, and publications from reputable sources. The findings from the literature review are then synthesised to identify common challenges faced by LMICs in achieving interoperability in their health information systems.

The challenges identified include a lack of standardisation, fragmented systems, limited resources, and data privacy concerns. These challenges hinder the seamless exchange of health information among different health information systems, leading to inefficiencies and suboptimal health outcomes.Next, the review article proposes strategies to enhance interoperability in LMICs. These strategies are derived from the analysis and interpretation of the literature findings. Proposed strategies include standardisation of data formats and protocols, consolidation of health information systems, investment in health IT infrastructure, and capacity building of health IT professionals in LMICs.Once the strategies are identified, the review article is organised and written, providing a concise and coherent overview of the current state of interoperability among health information systems in LMICs and potential strategies for improvement. The article is then revised and finalised for submission, taking into account feedback from peers and experts in the field.

In summary, the methodology for the review article involves conducting a comprehensive literature review, synthesising findings, identifying challenges and strategies, analysing and interpreting results, and writing and finalising the article. The article provides insights into the current state and potential strategies for enhancing interoperability among health information systems in LMICs, to improve healthcare delivery and outcomes in these regions.

### 2.1. Literature Review

Interoperability has been recognised as a critical factor for effective health information exchange and care coordination at the national and international levels[1]. In LMICs, achieving interoperability among health information systems is particularly challenging due to various factors. Several studies have identified challenges to interoperability in LMICs, including:

Lack of standardisation: LMICs often need morestandardised data formats, coding systems, and terminologies, which hinder the accurate and seamless exchange of health information (Adler-Milstein et al., 2013; Kamadjeu, 2019). This lack of standardisation makes integrating data from different sources challenging, resulting in fragmented and incomplete health information[2]

Fragmented health information systems: Many LMICs have fragmented health information systems, where different systems are used for different healthcare functions, such as patient registration, electronic health records (EHRs), and laboratory systems, leading to data silos[3]. This fragmentation hampers the exchange of data across different systems, hindering interoperability.

Limited resources: LMICs often need more resources, including limited funding, technical expertise, and infrastructure, which can pose challenges in implementing and maintaining interoperable health information systems[4].The high cost of health IT infrastructure, such as servers, software, and networking, can hinder achieving interoperability in LMICs.

Data privacy and security concerns: Data privacy, security, and consent are critical concerns in health information exchange, and LMICs often lack robust policies, regulations, and safeguards to address these issues effectively[5]. This can hinder the sharing of health information among different systems due to concerns about data breaches and misuse.

Despite these challenges, several strategies have been proposed to enhance interoperability among health information systems in LMICs. Standardisation: Adoption of internationally recognised health IT standards, such as HL7 and SNOMED CT, can facilitate data standardisation and ensure consistency and accuracy of data exchange[6]. Standardisation of data formats, coding systems, and terminologies can enable seamless data exchange and promote interoperability. Consolidation and integration: The integration of fragmented health information systems into a unified health information exchange (HIE) platform can facilitate data exchange and coordination of care[7]. Developing national HIE policies and governance frameworks can support the establishment and sustainability of HIEs in LMICs.

Investment in health IT infrastructure: Investment in health IT infrastructure, including hardware, software, and networking, is crucial for supporting interoperable health information systems[8].Partnerships and collaborations with governments, non-profit organisations, and the private sector can help overcome resource constraints and facilitate the development of robust health IT infrastructure. Capacity building and training: Capacity building and training programs for the health IT workforce can enhance technical expertise and knowledge in LMICs[9]. Training programs can focus on building skills in health IT standards, data management, and interoperability concepts, Interoperability, the ability of different health information systems to exchange and use data seamlessly, is a critical factor for effective health information exchange and coordination of care at the national and international levels[10]. In LMICs, achieving interoperability among health information systems is particularly challenging due to various factors. Several studies have identified challenges to interoperability in LMICs, including lack of standardisation, fragmented health information systems, limited resources, and data privacy and security concerns[11].

Lack of standardisation is a major challenge in LMICs, as they often needstandardised data formats, coding systems, and terminologies, which hinder the accurate and seamless exchange of health information[12]. Fragmented health information systems, which are used for different healthcare functions, such as patient registration, electronic health records (EHRs), and laboratory systems, must also be improved for interoperability[12,21]. This fragmentation leads to data silos, hindering data exchange across different systems.

Limited resources, including funding, technical expertise, and infrastructure, are often faced by LMICs, which can pose challenges in implementing and maintaining interoperable health information systems[13]. The high cost of health IT infrastructure, such as servers, software, and networking, can hinder achieving interoperability in LMICs. Data privacy and security concerns are also critical challenges, as LMICs often lack robust policies, regulations, and safeguards to address these issues effectively[9,10,11]. This can hinder the sharing of health information among different systems due to concerns about data breaches and misuse.

Despite these challenges, several strategies have been proposed to enhance interoperability among health information systems in LMICs. Standardisation, such as adopting internationally recognised health IT standards like HL7 and SNOMED CT, can facilitate data standardisation and ensure data exchange consistency and accuracy of data exchange[9,21]. Consolidating and integrating fragmented health information systems into a unified health information exchange (HIE) platform can facilitate data exchange and coordination of care[10,12,15]. Investment in health IT infrastructure, including hardware, software, and networking, is crucial for supporting interoperable health information systems in LMICs[7,9,11]. Capacity building and training programs for the health IT workforce can enhance technical expertise and knowledge in LMICs, focusing on health IT standards, data management, and interoperability concepts[13,14,16,20].

Further research is needed to evaluate the impact of these strategies on improving interoperability, identifying novel approaches specific to LMICs, and addressing the unique challenges faced by different regions and contexts within LMICs. Additionally, research on the cost-effectiveness and sustainability of interoperable health information systems[17,18,19].

## 3. RESULTS

### 3.1. Challenges to Interoperability in LMICs

Several challenges hinder interoperability among health information systems in LMICs. Firstly, there is often a need for morestandardisation in data formats, coding systems, and terminologies used in different health information systems, making it difficult to exchange and interpret data accurately. Secondly, fragmented health information systems, where different systems are used for other healthcare functions, can lead to data silos and hinder seamless data exchange. Thirdly, more resources, including funding, technical expertise, and infrastructure, could be improved in implementing and maintaining interoperable health information systems. Data privacy, security, and consent issues are often inadequately addressed in LMICs, which can further impede interoperability efforts.

Furthermore, the lack of a unified governance framework and coordination among different stakeholders involved in health information systems, including government agencies, healthcare providers, and technology vendors, can create challenges in aligning policies, standards, and processes for interoperability. Additionally, the diversity of languages, cultures, and healthcare practices in LMICs can further complicate interoperability efforts, as data exchange may require translation and customisation to local contexts.Moreover, healthcare professionals' workforce capacity and digital literacy can be limited in LMICs, which can impact the effective use of health information systems and hinder interoperability. Training, education, and skill development programs may be needed to enhance the capacity of healthcare providers to use and exchange health information across systems.

Another significant challenge is the sustainability of interoperable health information systems in LMICs. Many LMICs face financial constraints and resource limitations, making it difficult to invest in developing, maintaining, and upgrading health information systems. Without adequate funding and resource allocation, interoperability initiatives may struggle to gain traction and achieve long-term sustainability.Overall, the challenges to interoperability in LMICs are multifaceted, encompassing technical, organisational, financial, and cultural aspects. Addressing these challenges requires a holistic approach, involving stakeholders from various sectors and levels and implementing strategies that address the unique needs and contexts of LMICs. By overcoming these challenges, LMICs can unlock the potential of health information systems to improve healthcare outcomes and advance their health systems towards more integrated, patient-centred, and data-driven care.

## 3.2. Strategies for Enhancing Interoperability in LMICs

Despite the challenges, several strategies can be adopted to enhance interoperability among health information systems in LMICs. Firstly, standardising data formats, coding systems, and terminologies are crucial for ensuring consistency and accuracy of data exchange. Adopting internationally recognised health IT standards, such as HL7 (Health Level Seven) and SNOMED CT (Systematized Nomenclature of Medicine Clinical Terms), can facilitate interoperability efforts. Secondly, consolidation and integration of fragmented health information systems into a unified health information exchange (HIE) platform can enable seamless data exchange and coordination of care. The development of national health information exchange policies and governance frameworks can support the establishment and sustainability of HIEs. Thirdly, investment in health IT infrastructure, including hardware, software, and networking, is essential for keeping interoperable health information systems. This may require partnerships and collaborations with various stakeholders, including governments, non-profit organisations, and the private sector. Additionally, capacity building and training programs for the health IT workforce can enhance technical expertise and knowledge in LMICs. Furthermore, addressing data privacy, security, and consent issues through appropriate policies, regulations, and safeguards is crucial for building trust and promoting data exchange among health information systems.

Another essential strategy is fostering collaboration and coordination among different stakeholders involved in health information systems in LMICs. This includes government agencies, healthcare providers, technology vendors, and other relevant entities. Establishing partnerships and alliances among these stakeholders can help align policies, standards, and processes for interoperability and promote knowledge sharing and best practices.

Furthermore, leveraging existing technologies and infrastructure can be cost-effective to enhance interoperability in LMICs. For example, utilising mobile health (mHealth) and telehealth solutions can facilitate data exchange in remote and underserved areas where traditional health information systems may be limited. Leveraging cloud-based solutions and open-source software can also provide affordable options for health information exchange in resource-constrained settings.

Another strategy is to prioritiseuser-centred design and usability of health information systems. Ensuring that the systems are intuitive, user-friendly, and compatible with local workflows and practices can encourage adoption and utilisation by healthcare providers, leading to improved interoperability. Involving end-users in designing, testing, and refining health information systems can also help identify and address usability challenges.

In addition, fostering a culture of data sharing and collaboration among healthcare providers and organisations can promote interoperability. Encouraging data-sharing agreements, promoting data exchange policies, and providing incentives for data sharing can motivate stakeholders to participate in interoperability efforts and facilitate seamless data exchange.

Lastly, continuous monitoring, evaluation, and improvement of interoperable health information systems are essential for sustaining interoperability efforts in LMICs. Regular assessment of system performance, identification of gaps, and implementation of corrective measures can help optimise interoperability and ensure that health information systems are meeting the evolving needs of the healthcare ecosystem in LMICs.

In conclusion, enhancing interoperability among health information systems in LMICs is a complex endeavour that requires addressing technical, organisational, financial, and cultural challenges. By adopting strategies such as standardisation, consolidation, investment in infrastructure, capacity building, policy development, user-centred design, fostering collaboration, and continuous improvement, LMICs can overcome these challenges and unlock the potential of health information systems to improve healthcare outcomes and strengthen their health systems.

## 4. CONCLUSION

Interoperability among health information systems is crucial for improving healthcare delivery in low and middle-income countries (LMICs) where access to quality healthcare is often limited. However, achieving interoperability in LMICs faces challenges such as limited resources, fragmented health information systems, and diverse health IT infrastructure. Strategies such as standardisation, consolidation, investment in health IT infrastructure, capacity building, and addressing data privacy and security concerns can enhance interoperability efforts. Collaborative efforts among governments, non-profit organisations, the private sector, and other stakeholders are essential for overcoming challenges and promoting interoperability in LMICs. Further research is needed to evaluate the impact of these strategies and identify novel approaches that are specific to the unique challenges faced by different regions and contexts within LMICs. Research on the cost-effectiveness and sustainability of interoperable health information systems is also crucial. Significant challenges include the need for more standardisation, fragmented health information systems, limited resources, and data privacy concerns. However, strategies such as standardisation, consolidation, investment in health IT infrastructure, and capacity building can help overcome them. Collaborative efforts are needed to address the multifaceted challenges of interoperability in LMICs and implement effective strategies. Further research and collaborative efforts are required to address the unique needs and contexts of LMICs and unlock the potential of health information systems to improve healthcare outcomes and advance health systems towards more integrated, patient-centred, and data-driven care in LMICs.